\documentclass[sigconf]{acmart}

\AtBeginDocument{%
  \providecommand\BibTeX{{%
    \normalfont B\kern-0.5em{\scshape i\kern-0.25em b}\kern-0.8em\TeX}}}

\setcopyright{rightsretained}

\usepackage[T1]{fontenc} 
\usepackage[utf8]{inputenc}
\usepackage{todonotes}
\usepackage[hyphenbreaks]{breakurl}
\usepackage{fancyvrb}
\usepackage{pmboxdraw}
\usepackage{listings}
\usepackage{balance}

\graphicspath{{./figures/}}

\begin{document}

\title{Reproducibility Companion Paper: Describing Subjective Experiment
Consistency by $p$-Value P--P Plot}

\author{Jakub Nawała}
\email{jnawala@agh.edu.pl}
\orcid{0000-0002-5671-3726}
\author{Lucjan Janowski}
\orcid{0000-0002-3151-2944}
\affiliation{%
  \institution{AGH University of Science and Technology,
  Institute of Telecommunications}
  \city{Kraków}
  \country{Poland}
}

\author{Bogdan Ćmiel}
\affiliation{%
  \institution{AGH University of Science and Technology,
  Department of Mathematical Analysis, Computational Mathematics and Probability Methods}
  \city{Kraków}
  \country{Poland}
}

\author{Krzysztof Rusek}
\orcid{0000-0003-4336-7841}
\affiliation{%
  \institution{AGH University of Science and Technology,
  Institute of Telecommunications}
  \city{Kraków}
  \country{Poland}
}

\author{Marc A. Kastner}
\orcid{0000-0002-9193-5973}
\affiliation{%
  \institution{National Institute of Informatics}
  \city{Tokyo}
  \country{Japan}
}

\author{Jan Zahálka}
\orcid{0000-0002-6743-3607}
\affiliation{%
  \institution{Czech Technical University in Prague}
  \city{Prague}
  \country{Czech Republic}
}

\renewcommand{\shortauthors}{Nawała, Janowski, et al.}

\fancyhead{}

\begin{abstract}
In this paper we reproduce experimental results presented in our earlier
work titled ``Describing Subjective Experiment Consistency by
$p$-Value P--P Plot'' that was presented in the course of the 28th
ACM International Conference on Multimedia. The paper aims at
verifying the soundness of our prior results and helping others
understand our software framework. We present artifacts that
help reproduce tables, figures
and all the data derived from raw subjective responses that were
included in our earlier work. Using the artifacts we show
that our results are reproducible. We invite everyone
to use our software framework for subjective responses analyses
going beyond reproducibility efforts.
\end{abstract}




\copyrightyear{2021} 
\acmYear{2021} 
\acmConference[MM '21] {Proceedings of the 29th ACM International Conference on Multimedia}{October 20--24, 2021}{Virtual Event, China.}
\acmBooktitle{Proceedings of the 29th ACM International Conference on Multimedia (MM '21), October 20--24, 2021, Virtual Event, China}
\acmPrice{}
\acmDOI{10.1145/3474085.3477935} 
\acmISBN{978-1-4503-8651-7/21/10} 

\maketitle

\section{Artifacts Organisation}
\label{sec:artifacts_organisation}
\begin{sloppypar}
The artifacts are available for download from the following GitHub
repository: \burl{https://github.com/Qub3k/subjective-exp-consistency-check}
\cite{repo}.
Its file structure is presented in Fig.~\ref{fig:repo_structure}.
\end{sloppypar}

The \texttt{reproduce.py} script is the entry point, serving as a master script
governing the reproducibility process. The script's help message is designed so
as to provide sufficient information necessary to understand framework's
operation.
The following snippet shows how to invoke the
\texttt{reproduce.py}'s help
message using the command-line interface (CLI).
\begin{lstlisting}
$ python3 reproduce.py -h
\end{lstlisting}
Should the help message be insufficient, the reader is encouraged to take
a look at the \texttt{README.md} file. More specifically, its
``Reproducibility'' section provides further guidance on how to use
the framework. At last, since the framework is entirely open-sourced,
its operation can be investigated by looking at the source code.
\begin{figure}[h]
	\centering
\begin{Verbatim}
.
├── .gitignore
├── G_test_on_real_data.py
├── G_test_results.csv
├── LICENSE
├── README.md
├── _logger.py
├── bootstrap.py
├── figures/
├── friendly_gsd.py
├── gsd.py
├── gsd_prob_grid.pkl
├── hdtv1_exp1_scores_pp_plot_ready.csv
├── p_value_pp_plot_in-depth_tutorial.ipynb
├── probability_grid_estimation.py
├── qnormal.py
├── qnormal_prob_grid.pkl
├── reproduce.py
├── reproducibility/
├── requirements.txt
├── subjective_quality_datasets.csv
└── typical_vs_atypical.csv
\end{Verbatim}
	\caption{Structure of the GitHub repository.}
	\label{fig:repo_structure}
\end{figure}

Another two important files in the repo are: (i)
\texttt{sub\-jec\-tive\_qua\-li\-ty\_da\-ta\-sets.csv}
and (ii) \texttt{G\_test\_results.csv}. The former one includes raw subjective data
that is processed in the original paper \cite{Nawala2020ACM}. The most important
output of this processing is the \texttt{G\_test\_re\-sults.csv} file. It includes
results of running computationally intensive bootstrapped version of the
G-test of goodness-of-fit (cf. Fig. 2 from \cite{Nawala2020ACM}).
Effectively, recreating these results is the
most significant part of the reproducibility efforts.

\section{Setup and Execution}
\label{sec:setup_and_execution}
We recommend to create a separate Python virtual environment and install there
all the dependencies listed in the \texttt{re\-quire\-ments.txt} file. Importantly,
Python version 3.7 or newer is required.
When the required packages are installed the \texttt{reproduce.py} script
can be run through the
CLI, as shown below.
\begin{lstlisting}
$ python3 reproduce.py [-h] [-n N] {scenario}
\end{lstlisting}
The \texttt{\{sce\-nar\-i\-o\}}
place-holder identifies the execution scenario. As of the time of writing
this paper there are five such scenarios. They are identified by subsequent
integers from the range 1--5. The following list provides details on each
scenario.
\begin{enumerate}
	\item Reproduce the original experiments using existing G-test results
	(i.e., the \texttt{G\_test\_re\-sults.csv} file). This executes immediately,
	but in principle does not reproduce the most important piece of the data
	presented in the original paper.
	\item Reproduce only these G-test results that are necessary to draw
	Fig.~3 from the original paper. This already takes a significant amount
	of time to run (approximately nine days according to our internal tests).
	This scenario makes sense since Fig.~3 is a central part of the discussion
	presented in the original paper.
	\item Reproduce all G-test results. This scenario reproduces from
	the scratch all the results presented in the original paper. As such
	this is the most important scenario of the reproducibility framework.
	However, according to our internal tests, this takes around 21 days to run.
	\item Run the G-test for randomly selected \texttt{N} stimuli (cf. the
	\texttt{-n N} argument of the call to \texttt{re\-pro\-duce.py} above).
	This scenario can be used to quickly
	check the correctness of \texttt{N} randomly selected results from the
	\texttt{G\_test\_re\-sults.csv} file. We note that according to our internal
	tests processing one stimulus takes about four to seven minutes.
	\item Reproduce the probability grids of the Generalised Score Distribution
	(GSD) and Qunatized Normal (QNormal) models (cf. Sec. GSD Parameters
	Estimation in \cite{Nawala2020ACM}). Those grids are internally used when
	running the G-test.
\end{enumerate}

Since our framework is implemented in Python in its entirety, it can
be run on any platform.\footnote{We confirmed framework's operation on the
three popular operating systems: Windows 10, Mac OS 10.15 and Ubuntu Linux 18.04.05.}
All execution times mentioned in this paper were measured using the following
hardware setup: Intel Core i3-8130U CPU, 16 GB of 2400 MHz RAM and 256 GB SSD disk
(Lenovo LENSE30256GMSP34MEAT3TA).

\subsection{Batch Processing Capability}
\label{ssec:batch_processing_capability}
Reproducing complete G-test results takes a significant amount of time
when done on a single machine. Thus, we make available the batch
processing friendly variation of the G-test running framework. It can be
used to run multiple parallel instances of the G-test, each running on
a different chunk of the input data.
Crucially, the \texttt{reproduce.py}
script does not support batch processing, as this would greatly complicate
its structure. Instead, the \texttt{friendly\_gsd.py} script must be used.
Still, both \texttt{reproduce.py} and \texttt{friendly\_gsd.py} scripts
will produce the same output.
For more information we refer the reader to the ``Batch Processing''
section of the \texttt{README.md} file.

\section{Experiments}
\label{sec:experiments}
Execution scenarios 1, 2 and 3 (cf. Sec~\ref{sec:setup_and_execution})
reproduce all tables and figures presented in the original paper. However,
only scenario 3 reproduces entirely the data used to generate figures and
tables. The other two scenarios either use the data processing outputs
from the original paper (scenario 1) or reproduce only a part of the data
(scenario 2).

Since the G-test running framework internally uses pre-cal\-cu\-lat\-ed probability
grid of the GSD model, to achieve the complete reproducibility (i.e.,
being able to achieve the same results when being provided only with
raw subjective data contained in the
\texttt{sub\-jec\-tive\_qua\-li\-ty\_da\-ta\-sets.csv} file) one has to run
execution scenario 5 as well.

All in all, both scenario 3 and scenario 5 must be executed to check
results reproducibility.
The snippet below shows two
calls to the \texttt{re\-pro\-duce.py}
script that fulfil this goal.
\begin{lstlisting}
$ python3 reproduce.py 3
$ python3 reproduce.py 5
\end{lstlisting}
The first call produces four types of
output: (i) CSV files with reproduced tables contents, (ii) PDF files
with reproduced figures, (iii) a CSV file with G-test results and (iv)
tables contents written to the standard output. Fig.~\ref{fig:scenario_three_output}
shows files that are created as a result of this call. Significantly,
the reproduced figures have the same
formatting as the one used in the original paper (cf. Fig.~\ref{fig:ex_repro_fig}).

We also note here that the G-test used in the framework has randomness
built into it. Thus, reproduced results will not exactly match the
ones generated for the purposes of the original paper. This is because
we use the bootstrapped version of the G-test that internally generates
10,000 synthetic random samples based on each observed sample. We
refer readers interested into more details on the topic
to section ``In-depth Tutorial about
Generating $p$-Value P--P Plots for Your Subjective Data'' of the
\texttt{README.md} file in the GitHub repository.

\begin{figure}
	\centering
\begin{Verbatim}
p-value_pp-plot_fig_one_a.pdf
p-value_pp-plot_fig_one_b.pdf
table_one_score_distribution.csv
table_two_pvals.csv
p-value_pp-plot_HDTV1_fig_three.pdf
p-value_pp-plot_ITS4S2_fig_three.pdf
p-value_pp-plot_ITS4S_AGH_fig_three.pdf
p-value_pp-plot_AGH_NTIA_fig_three.pdf
table_three_five_lowest_pvalue_res_its4s_agh.csv
table_four_five_lowest_pvalue_res_its4s2.csv
G_test_on_subjective_quality_datasets_chunk000_of_001.csv
\end{Verbatim}
	\caption{Files generated as a result of running execution scenario 3.}
	\label{fig:scenario_three_output}
\end{figure}
\begin{figure}
	\centering
	\includegraphics[width=\columnwidth]{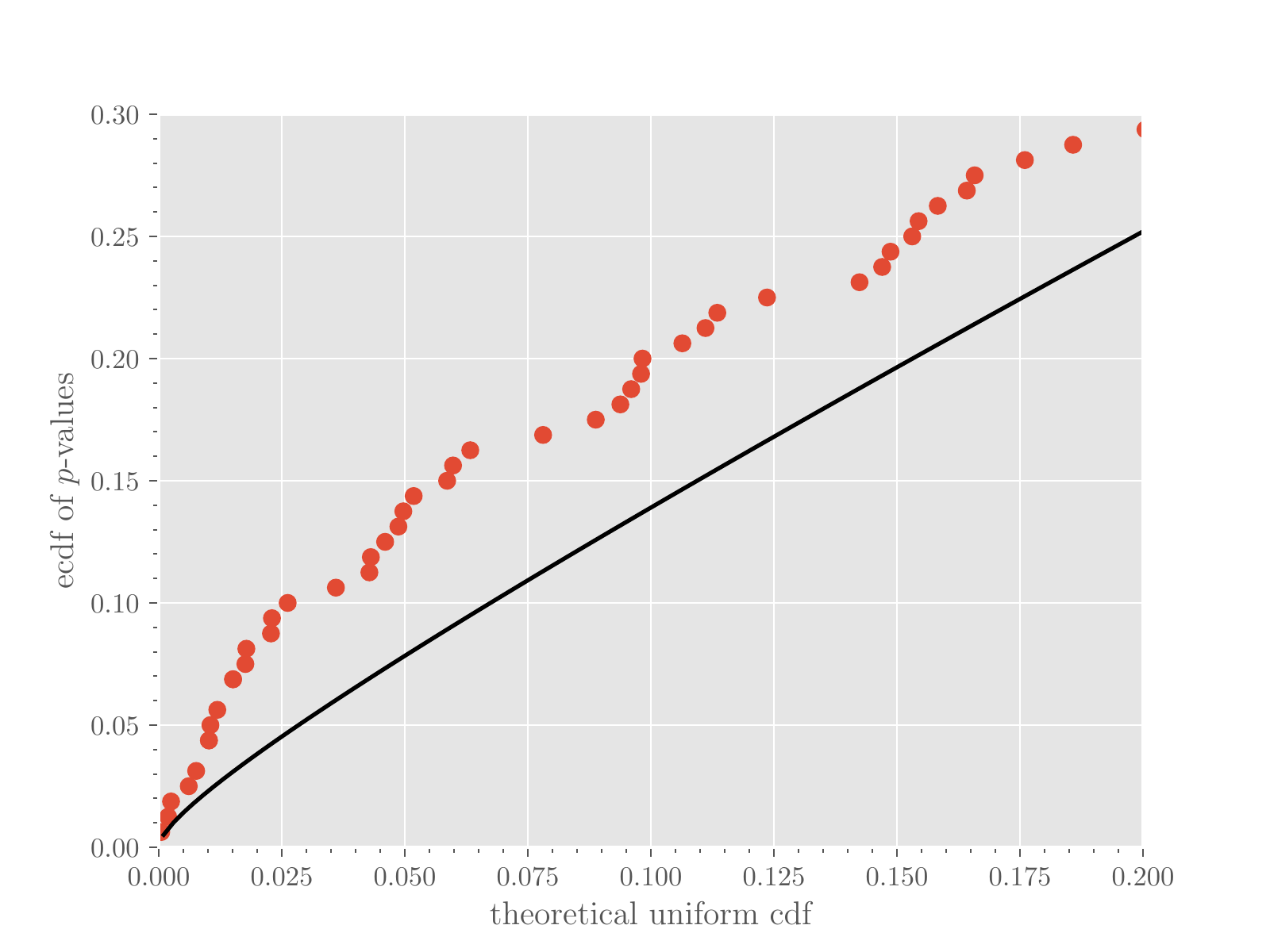}
	\caption{Reproduced Fig. 1b from the original paper. Please note that
	the formatting of the figure is the same as in the original paper.}
	\label{fig:ex_repro_fig}
\end{figure}

Running scenario 5 (the second call from the snippet above)
creates two files only: (i) \texttt{re\-pro\-duced\_gsd\_prob\-\_grid.pkl}
and (ii) \texttt{re\-pro\-duced\_qnormal\_prob\-\_grid.pkl}.
They are pickled Python objects and more specifically,
pickled Pandas DataFrames. They can be manually compared with the
corresponding pickle files from the original paper: \texttt{gsd\_prob\_grid.pkl}
and \texttt{qnor\-mal\_prob\-\_grid.pkl}.

We note that our framework reproduces Fig. 1a and 1b from the original
paper using ready-made CSV files.\footnote{The CSV files are available in
the \texttt{reproducibility} folder in the GitHub repository.} This is
because the two plots use synthetic data. Differently put, the plots
were not generated from subjective responses gathered during any
real-life subjective experiment.

\vfill\eject
\section{Reproducibility Efforts}
\label{sec:reproducibility}

The code is open-source, well readable, and sufficiently commented. All results of the original paper are easily reproducible, directly generating the figures used in the original paper. 

Since the submitted software was of high quality in the first version already, the reproducibility review has mostly consisted of minor fixes and ease-of-use improvements. Firstly, the review process resulted in the batch processing mode of the software being more accessible to the user. This is essential, as the sequential mode runs for days to reproduce all results. Secondly, the authors have fixed the random stimuli scenario that did not fail gracefully when the \textsc{-n} parameter was omitted (now the default is 3 stimuli). Finally, the authors have been very responsive not only to the reviewer comments, but also to general GitHub user comments. 

All of the above aspects lead us to believe this is a software worthy of the reproducibility badge.

\section{Invitation}
\label{sec:invitation}
Although this paper focuses on reproducibility, our GitHub repository
\cite{repo} was created to help others use our framework in
future analyses as well. We invite everyone, who has at hand
a data set of subjective responses, to use the framework. It can test
how well the GSD models subjective responses distribution and provide
insights into subjective experiment consistency. For more details
we refer the reader to our original paper \cite{Nawala2020ACM} and
the \texttt{README.md} file in the repository.

\begin{acks}
This work was supported by
the Polish Ministry of Science and Higher Education
with the subvention funds of
the Faculty of Computer Science,
Electronics and Telecommunications of AGH University
and by the PL-Grid Infrastructure. Furthermore, the research leading
to these results has received funding from the Norwegian Financial
Mechanism 2014--2021 under project 2019/34/H/ST6/00599.

We also kindly acknowledge the help of Franz Hahn, who identified
and helped us resolve an issue with our source code.
\end{acks}

\bibliographystyle{ACM-Reference-Format}
\bibliography{bibliography}


\end{document}